\documentclass[aps,prc,preprint,showpacs,superscriptaddress]{revtex4}  
\usepackage{graphicx}
\usepackage{amsmath}
\usepackage{xcolor}
\newcommand {\nc} {\newcommand}
\nc {\beq} {\begin{eqnarray}} \nc {\eol} {\nonumber \\} \nc {\eeq}
{\end{eqnarray}} \nc {\eeqn} [1] {\label{#1} \end{eqnarray}} \nc
{\eoln} [1] {\label{#1} \\} \nc {\ve} [1] {\mbox{\boldmath $#1$}}
\nc {\rref} [1] {(\ref{#1})} \nc {\Eq} [1] {Eq.~(\ref{#1})} \nc
{\la} {\mbox{$\langle$}} \nc {\ra} {\mbox{$\rangle$}} \nc {\dd}
{\mbox{${\rm d}$}} \nc {\cM} {\mathcal{M}} \nc {\cY} {\mathcal{Y}}
\nc {\dem} {\mbox{$\frac{1}{2}$}} \nc {\ut} {\mbox{$\frac{1}{3}$}}
\nc {\qt} {\mbox{$\frac{4}{3}$}} \nc {\Li} {\mbox{$^6\mathrm{Li}$}}
\nc {\M} {\mbox{$\mathcal{M}$}}  \nc {\arrow} [2]
{\mbox{$\mathop{\rightarrow}\limits_{#1 \rightarrow #2}$}}

\begin{document}

\title{Analysis of the  $^{3}{\rm He}(\alpha, \gamma)^{7}{\rm Be}$ and
$^{3}{\rm H}(\alpha,\gamma)^{7}{\rm Li}$  astrophysical direct
capture reactions in a modified potential-model approach}

\author{E.M. Tursunov}
\email{tursune@inp.uz} \affiliation{Institute of Nuclear Physics,
Academy of Sciences, 100214, Ulugbek, Tashkent, Uzbekistan}

\author{S.A. Turakulov}
\email{turakulov@inp.uz} \affiliation{Institute of Nuclear Physics,
Academy of Sciences, 100214, Ulugbek, Tashkent, Uzbekistan}

\author{A.S. Kadyrov}
\email{a.kadyrov@curtin.edu.au} \affiliation{Curtin Institute for
Computation and Department of Physics and Astronomy, Curtin
University, GPO Box U1987, Perth, WA 6845, Australia}

\begin{abstract}
Astrophysical $S$ factors and reaction rates of the direct
radiative capture processes $^{3}{\rm He}(\alpha, \gamma)^{7}{\rm
Be}$ and $^{3}{\rm H}(\alpha,\gamma)^{7}{\rm Li}$, as well as the
primordial abundance of the $^{7}{\rm Li}$ element, are estimated in
the framework of a modified two-body potential model. It is shown
that suitable modification of phase-equivalent
$\alpha-^{3}${\rm He} potentials in the $d$ waves can improve the
description of the astrophysical $S$ factor for the direct $^{3}{\rm
He}(\alpha, \gamma)^{7}{\rm Be}$ radiative capture reaction at energies above 0.5 MeV. An estimated $^{7}{\rm Li/H}$
abundance ratio of $
(4.89\pm 0.18 )\times 10^{-10}$
is in very good agreement with the recent measurement of $
(5.0\pm 0.3) \times 10^{-10}$ by the LUNA collaboration.

\end{abstract}

\maketitle

\section{Introduction}

 Realistic estimation of the primordial abundances of the lithium isotopes $^{6}{\rm Li}$ and $^{7}{\rm Li}$,
  the two heaviest elements in Big Bang nucleosynthesis (BBN),
 is one of the most important and unsolved problems of nuclear astrophysics.
 The primordial abundances of these elements can be extracted from an analysis of astronomical
  observations of old metal-poor halo stars. For the $^{7}{\rm Li}$ abundance astronomical data provide
   a value of $^{7}{\rm Li}/{\rm H}$=(1.58$^{+0.35}_{-0.28})\times 10^{-10}$  \cite{sbor10} which is 2 to 4 times less than an
   estimate of $^{7}{\rm Li}/{\rm H}$=(4.68$\pm $0.67)$\times 10^{-10}$  of the BBN model  \cite{cyb16}.
   On the other hand, there is the so-called second lithium problem which is related to the abundance ratios of the lithium isotopes.
   A recent analysis of the direct measurements data of the LUNA collaboration yielded a value $^{6}{\rm Li} /^{7}{\rm Li}= (1.6\pm 0.3)\times 10^{-5}$
    \cite{luna17} which is 3 orders of magnitude lower than the astronomical observation \cite{asp06}.
   These problems were subjects of intense discussions during a recent topical workshop \cite{WS2019}. 
   This demonstrates that they are still far from being solved.

An important question is whether or not the lithium problems originate from astronomy or nuclear physics.
From one side a small primordial abundance of the $^{6}{\rm Li}$ element is well described in nuclear physics from both experimental and theoretical perspectives. This element was mainly produced during the BBN epoch via the direct capture
$d(\alpha, \gamma)^{6}{\rm Li}$ process. Until recently the main problem in theoretical studies of this process was connected with a
 consistent description of  the isospin-forbidden E1 transition. Finally, results of theoretical calculations within the most realistic three-body
  model \cite{tur16,bt18,tur18,tur20} are now in very good agreement with the direct data of the LUNA collaboration \cite{luna14,luna17}.
Good agreement was obtained for all observable of practical interest including astrophysical $S$ factor, reaction rates and the primordial abundance of the $^{6}{\rm Li}$ element.
 The absolute values and temperature dependence of the reaction rates of the LUNA data have been reproduced with a good accuracy, which was
 a consequence of the correct treatment of the isospin-forbidden E1 transition in contrast to two-body models based on so-called
  exact-mass prescription \cite{tur15,tur20}.
 The calculated value of $(0.67 \pm 0.01) \times 10^{-14}$  \cite{tur18,tur20} for the $^6$Li/H primordial abundance ratio is consistent
  with the estimate $(0.80 \pm 0.18) \times 10^{-14}$ of the LUNA collaboration \cite{luna17}.

The  $^{7}{\rm Li}$ isotope was produced mainly through radiative capture reactions $^{3}{\rm He}(\alpha,
\gamma)^{7}{\rm Be}$ and $^{3}{\rm H}(\alpha, \gamma)^{7}{\rm Li}$
during the BBN period  \cite{adel11}. These direct capture reactions play a significant role also in stellar nucleosynthesis \cite{fields11},
as well as in the pp chain of solar hydrogen
burning \cite{serene13}. The primordial abundance of the  $^{7}{\rm Li}$ element
is evaluated from the reaction rates of the two capture processes mentioned above.
The $^{7}{\rm Be}$  nucleus is produced in the $^{3}{\rm He}(\alpha,
\gamma)^{7}{\rm Be}$ direct capture process and subsequently decays through electron
capture resulting in the $^{7}{\rm Li}$ element. The
  $^{3}{\rm H}(\alpha, \gamma)^{7}{\rm Li}$ process then gives a
small additional contribution to the lithium primordial abundance.

In recent years, the lithium abundance problem was discussed
extensively from both experimental and theoretical viewpoints
\cite{coc17,cyb16}. One has to note that experimental measurements of these
reactions in low-energy region face serious difficulties due to
strong Coulomb repulsion. Nevertheless, direct data for the astrophysical
$S$ factor of the  $^{3}{\rm He}(\alpha, \gamma)^{7}{\rm Be}$  capture
process at several energies around 100 keV were obtained  by
the LUNA collaboration in the underground facility
\cite{luna06,luna07}.
Later, this data set was
supplemented with a more accurate value of the astrophysical
$S$ factor at Gamow peak energy region, $S_{34}$(23$^{+6}_{-5}$
keV)=0.548 $\pm$ 0.054 keV, determined on the basis of observed neutrino fluxes
from the Sun within the standard solar model \cite{takacs15}. Based
on those results, the authors of Ref. \cite{takacs15} extracted an
estimate of 5.0$\times 10^{-10}$ for the $^{7}{\rm Li} /{\rm H}$
abundance, close to the standard BBN value and more than three times
larger than the astronomical data. Recently, the astrophysical
$S$ factor was reevaluated at the solar Gamov energy peak and its
value,
 $S_{34}$(23$^{+6}_{-5}$ keV)=0.590 $\pm$ 0.050 keV b, overlaps with the previous estimate within the error bars \cite{takacs18}.
Additionally, the data set for the reaction $^{3}{\rm He}(\alpha,
\gamma)^{7}{\rm Be}$ was recently extended up to 4.5 MeV in the
center-of-mass frame energy \cite{sz13,sz19}.

Theoretically, the astrophysical capture processes $^{3}{\rm
He}(\alpha, \gamma)^{7}{\rm Be}$ and $^{3}{\rm H}(\alpha,
\gamma)^{7}{\rm Li}$ have been studied in potential
\cite{mohr09,mason09,dub10,tur18_7Be} and microscopic models
\cite{kajino87,vasil12,sol14,sol19},  a microscopic R-matrix approach
\cite{desc10}, as well as in a semimicroscopic phenomenological
approach \cite{noll01}, a fermionic molecular dynamics (FMD) method
\cite{neff11} and a no-core shell model with continuum (NCSMC)
\cite{doh16,vor19}. The most realistic microscopic approaches
\cite{doh16,vor19,neff11,sol19} still have problems with
simultaneous description of the above mirror capture reactions,
including the both absolute values and energy dependence of the
astrophysical $S$ factor.

In Ref. \cite{tur18_7Be} a realistic potential model was developed
for the description of the capture reactions mentioned above. It was shown that
the potential model is able to describe the astrophysical $S$ factors
at low energies, below 0.5 MeV, which include the BBN energy region
of $E_{\rm cm}$=180-400 keV, leading to good agreement with the
experimental data \cite{luna06,luna07,takacs15}. However, the
existing data sets at intermediate energies are underestimated and
this discrepancy increases with the energy. An important question
is, whether the potential model can describe the astrophysical
$S$ factor of the direct capture processes $^{3}{\rm He}(\alpha,
\gamma)^{7}{\rm Be}$ and $^{3}{\rm H}(\alpha, \gamma)^{7}{\rm Li}$
at low and intermediate energies simultaneously. Answering this question
may have important implications for both nuclear theory and astrophysical applications.

The aim of the present study is to describe the existing data for the astrophysical $S$ factors of the
 $^{3}{\rm He}(\alpha, \gamma)^{7}{\rm Be}$ and $^{3}{\rm H}(\alpha,
\gamma)^{7}{\rm Li}$ direct capture reactions at low- and
intermediate-energy regions and to estimate the reaction rates of
these processes and the primordial abundance of the $^{7}{\rm Li}$
element in the potential model. As it is known from the literature
\cite{tur18_7Be}, the dipole E1-transition operator yields the main
contribution to the above processes at low and intermediate energies.
The E2 transition contributes only in the resonance energy region
near 3 MeV in the center-of-mass frame. The M1 transition is even
more suppressed and this is the case at all energies.

As it was shown in Ref. \cite{tur18_7Be}, below 0.5 MeV the main contribution to the E1 $S$ factor comes from the initial
$\alpha+^{3}{\rm He}$ and $\alpha+^{3}{\rm H}$ $s$-wave scattering states. However, at intermediate energies the role of the
 $d$-wave scattering states increases and their contribution becomes dominant beyond 2 MeV.
 On this basis it would be very useful to search for optimal $d$-wave $\alpha+^{3}{\rm He}$ and $\alpha+^{3}{\rm H}$  potentials,
 which would allow to better describe the astrophysical $S$ factor data for the aforementioned capture reactions.
 In this way we perform an optimization procedure among phase equivalent $\alpha+^{3}{\rm He}$ potentials in the partial
 $d_{3/2}$ and $d_{5/2}$ waves.

 The two-body Gaussian potentials \cite{dub10} will be
examined. In Ref. \cite{tur18_7Be} the potential parameters in the
$s$ wave were adjusted to reproduce the astrophysical $S$ factor
of the $\alpha+^{3}{\rm He}$ direct capture reaction at low energies
in addition to the phase shift data. In the $p_{3/2}$ and
$p_{1/2}$ partial waves the potential parameters were additionally
adjusted to reproduce the bound state properties: binding
energies and the values of the asymptotic normalization
coefficients (ANC) for the $^7$Be$(3/2^-)$ ground and $^7$Be$(3/2^-)$
excited states extracted from the analysis of the experimental data
within the DWBA method \cite{olim16}.

This article is organized as follows. In Section II the theoretical model will be
briefly described, Section III is devoted to the analysis of numerical results.
Conclusions will be drawn in the last section.

\section{Theoretical model}

Astrophysical $S$ factor of the radiation capture process is
expressed in terms of the cross section as \cite{nacre99}
\begin{eqnarray}
S(E)=E \, \, \sigma(E) \exp(2 \pi \eta)
\end{eqnarray}
where $E$ is the collision energy in the center-of-mass (cm) frame and $\eta$ is the
Sommerfeld parameter.
The cross section reads as \cite{nacre99,dub10}
\begin{eqnarray}
\sigma(E)=\sum_{J_f \lambda \Omega}\sigma_{J_f \lambda}(\Omega),
\end{eqnarray}
where $\Omega=$ E  or M (electric or magnetic transition), $\lambda$
is a multiplicity of the transition, $J_f$ is the total angular
momentum of the final state. For a particular final state with total
momentum $J_f$ and multiplicity $\lambda$ we have
\begin{align}
 \sigma_{J_f \lambda}(\Omega) =& \sum_{J}\frac{(2J_f+1)} {\left
[S_1 \right]\left[S_2\right]} \frac{32 \pi^2 (\lambda+1)}{\hbar
\lambda \left( \left[ \lambda
\right]!! \right)^2} k_{\gamma}^{2 \lambda+1} C^2(S) \nonumber \\
&\times \sum_{l S}
 \frac{1}{ k_i^2 v_i}\mid
 \langle \Psi_{l_f S}^{J_f}
\|M_\lambda^\Omega\| \Psi_{l S}^{J} \rangle \mid^2,
\end{align}
where $\Psi_{l S}^{J}$ and $\Psi_{l_f S}^{J_f}$ are the initial and final state
wave functions, respectively,
$M_\lambda^\Omega$ is the electric or magnetic transition operator,
$l,l_{f}$ are the orbital momenta of the initial and final
states, respectively, $k_i$ and $v_i$ are the wave number and
velocity of the $\alpha-^3$He (or $\alpha-^3$H) relative motion of
the entrance channel, respectively; $S_1$, $S_2$ are spins of the
clusters $\alpha$ and $^3${\rm He} (or $^3$H), $k_{\gamma}=E_\gamma
/ \hbar c$ is the wave number of the photon corresponding to energy
$E_\gamma=E_{\rm th}+E$, where $E_{\rm th}$ is the threshold energy.
The spectroscopic factor \cite{nacre99} $C^2(S)$  within the
potential approach is equal to 1, since the potential reproduces the
two-body experimental data, energies and phase shifts in partial
waves \cite{mukh16}. We also use short-hand notations $[S]=2S+1$ and
$[\lambda]!!=(2\lambda+1)!!$. Further details of the wave functions and
matrix element calculations can be found in Ref. \cite{tur18_7Be}.

\section{Numerical results}
\subsection{Details of the calculations and phase-shift descriptions}

We use simple Gaussian-form potentials for the $\alpha-^3$He and
$\alpha-^3$H two-body interactions \cite{dub10,tur18_7Be}:
 \begin{eqnarray}
 V^{lSJ}(r)=V_0 \exp(-\alpha_{0} r^2)+V_c(r),
\label{pot}
 \end{eqnarray}
 where the Coulomb part is given as
\begin{eqnarray}
 V_c(r)=
\left\{
\begin{array}{lc}
 Z_1 Z_2 e^2/r &  {\rm if} \,\, r>R_c, \\
Z_1 Z_2 e^2 \left(3-{r^2}/{R_c^2}\right)/(2R_c) &  {\rm otherwise},
\end{array}
\right. \label{Coulomb}
\end{eqnarray}
with the Coulomb parameter $R_c$, and charge numbers $Z_1$, $Z_2$ of
the first and second clusters, respectively. The parameters
$\alpha_0$ and $V_0$ of the potential are specified for each partial
wave. In Ref. \cite{tur18_7Be} we examined several potential
models for the description of the $\alpha-^3$He and $\alpha-^3$H
interactions. As discussed in the introduction, the d-wave
potentials can be further improved by modifying the depth ($V_0$)
and width ($\alpha_0$) parameters for the better description of the
astrophysical $S$ factors at intermediate energies.

\par The Schr\"{o}dinger equation in the entrance
and exit channels are solved with the
$\alpha-^{3}${\rm He} and $\alpha-^{3}$H central potentials as
defined in Eq.(\ref{pot}) with the corresponding Coulomb part from
Eq.(\ref{Coulomb}).
The same entry parameter values as in Ref. \cite{tur18_7Be} are
used: $\hbar^2/2m_{N}$=20.7343 MeV fm$^2$ and $R_c$=3.095 fm
(Coulomb parameter), however  the nuclear masses are taken as
$m_{^4{\rm He}}
=4 m_N$ and $m_{^3{\rm He}}=m_{^3{\rm H}}
=3 m_N$, where $m_N$ is the nucleon mass.

The expressions for the astrophysical $S$ factor and cross
section given above are valid only for the radial scattering wave function (the radial component of the initial state wave function $\Psi_{l S}^{J}$)
normalized at large distances as
\begin{eqnarray}
u_E^{(lSJ)}(r)\arrow{r}{\infty} \cos\delta_{lSJ}(E) F_l (\eta,kr) +
\sin\delta_{lSJ}(E) G_l(\eta,kr), \label{eq220}
\end{eqnarray}
where $k$ is the wave number of the relative motion,
$F_l$ and $G_l$ are regular and irregular
Coulomb functions, respectively, and $\delta_{lSJ}(E)$ is the phase
shift in the $(l,S,J)$th partial wave. The scattering wave function
$u_{E}(r)$ of the relative motion is calculated as a solution of the
Schr{\"o}dinger equation using the Numerov method with an
appropriate potential subject to the boundary condition specified in
Eq. (\ref{eq220}).

The depth and width parameters of the $\alpha-^{3}$He and
$\alpha-^{3}$H model potentials $V_D^n$ and $V_{M1}^n$ are given in
Tables \ref{tab1} and \ref{tab1a}, respectively. In 3th and 4th
columns of the tables the energies of forbidden states are
presented. The potentials contain two forbidden states in the $s$
waves, while a single forbidden states in the each of $p_{3/2}$,
$p_{1/2}$, $d_{3/2}$, $d_{5/2}$ partial waves. These potentials
differ from each other only in the $s$ and $p$ waves. At the same
time, model potentials $V_D^n$ and $V_{M1}^n$ are similar to
potentials $V_D^a$ and $V_{M1}^a$ from Ref. \cite{tur18_7Be},
respectively.
The only difference is in the $d$-wave parameter values. The latter
have now been fitted to better reproduce the astrophysical $S$
factors at larger energies.

\begin{table}[htb]
\caption{Values of the depth ($V_0$) and width ($\alpha_0$)
parameters of the $\alpha - ^3${\rm He} ($^3$H) potential $V_D^n$ in
different partial waves (see  Eq. (\ref{pot})).
}
{\begin{tabular}{@{}ccccc@{}} \toprule $L_J$ & \,\,\,\,\, $V_0$ (MeV) &
\,\,\,\,\, $\alpha_0$ (fm$^{-2}$) & \,\,\,\,\, E$^{^7\rm Be}_{\textrm{FS}}$ (MeV)&\,\,\,\,\,
E$^{^7\rm Li}_{\textrm{FS}}$ (MeV)
\\\colrule
$s_{1/2}$ & -78.0 & 0.186 & -40.03; -7.03 & -41.34; -8.09 \\
$p_{3/2}$ & -83.8065 & 0.15747 & -27.11 & -28.33 \\
$p_{1/2}$ & -82.0237 & 0.15747 & -26.02 & -27.24 \\
$d_{3/2}$ & -180.0 & 0.4173 & -11.96 & -13.22 \\
$d_{5/2}$ & -190.0 & 0.4017 & -18.13 & -19.39 \\
$f_{5/2}$ & -75.9 & 0.15747 & - & - \\
$f_{7/2}$ & -85.2 & 0.15747 & - & - \\\botrule
\end{tabular}\label{tab1}}
\end{table}

\begin{table}[htb]
\caption{Values of the depth ($V_0$) and width ($\alpha_0$)
parameters of the $\alpha - ^3${\rm He} ($^3$H) potential $V_{M1}^n$
in different partial waves (see  Eq. (\ref{pot})).
}
{\begin{tabular}{@{}ccccc@{}} \toprule $L_J$ & \,\,\,\,\, $V_0$ (MeV) &\,\,\,\,\,
$\alpha_0$ (fm$^{-2}$) & \,\,\,\,\, E$^{^7\rm Be}_{\textrm{FS}}$ (MeV)&\,\,\,\,\,
E$^{^7\rm Li}_{\textrm{FS}}$ (MeV)
\\\colrule
$s_{1/2}$ & -50.0 & 0.109 & -25.70; -5.17 & -26.95; -6.11 \\
$p_{3/2}$ & -75.59760 & 0.13974 & -24.58 & -25.78 \\
$p_{1/2}$ & -70.75751 & 0.13308 & -22.55 & -23.74 \\
$d_{3/2}$ & -180.0 & 0.4173 & -11.96 & -13.22 \\
$d_{5/2}$ & -190.0 & 0.4017 & -18.13 & -19.39 \\
$f_{5/2}$ & -75.9 & 0.15747 & - & - \\
$f_{7/2}$ & -85.2 & 0.15747 & - & - \\\botrule
\end{tabular}\label{tab1a}}
\end{table}

\begin{figure}[htb]
\includegraphics[width=16.8cm]{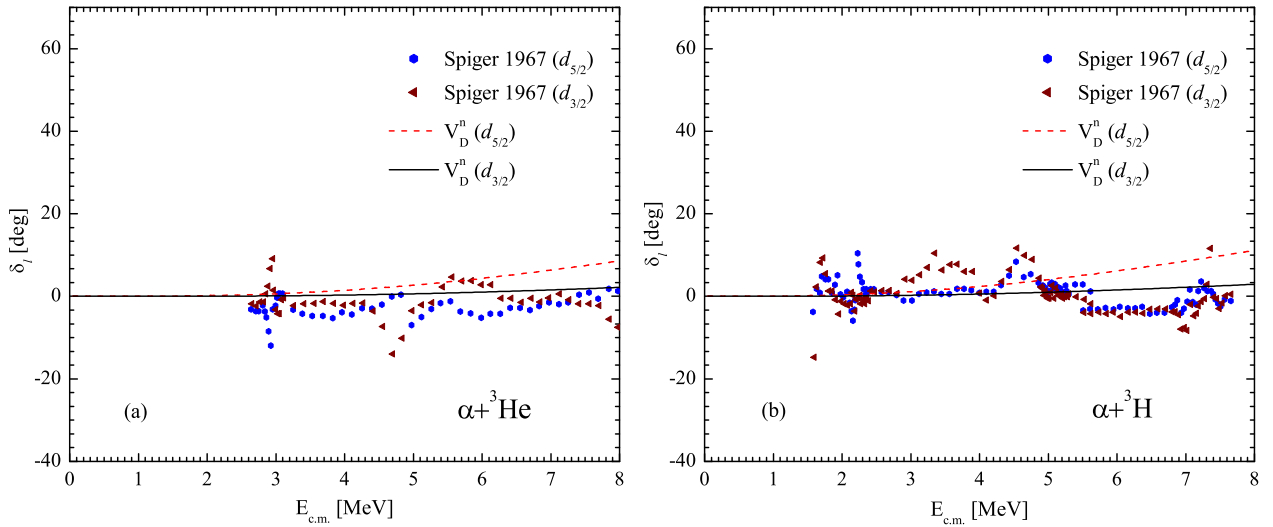}
\caption{$d$-wave phase shifts for the $^{3}{\rm
He}+\alpha$ (panel a) and $^{3}{\rm H}+\alpha$ (panel b) scattering within
potential models $V_D^n$ and $V_{M1}^n$  in comparison with experimental data from Ref. \cite{spiger}.} \label{f1}
\end{figure}

In Fig. \ref{f1} the experimental data \cite{spiger} for the
$^{3}{\rm He}+\alpha$ (panel a) and $^{3}{\rm H}+\alpha$ (panel b)  $d$-wave
scattering phase shift are compared with the theoretical calculations using the new model potentials
$V_D^n$ and $V_{M1}^n$. The phase shift description in the other partial
waves were given in Ref. \cite{tur18_7Be}. Additionally, the
presented models reproduce the energy spectrum of the $^{7}{\rm Be}$
and $^{7}{\rm Li}$ nuclei, as well as the empirical values of
the ANC for the ground $p_{3/2}$ and
the first excited $p_{1/2}$ bound states of the $^7$Be nucleus
\cite{tur18_7Be}. Indeed, the $V_D^n$ model yields $C(3/2^-)$=4.34
fm$^{-1/2}$ and $C(1/2^-)$=3.71 fm$^{-1/2}$, while the alternative
$V_{M1}^n$ model reproduces the ANC values of $C(3/2^-)$= 4.785 fm$^{-1/2}$ and
$C(1/2^-)$=4.242 fm$^{-1/2}$ extracted from the analysis of the experimental data using the DWBA method \cite{olim16}.

\subsection{Astrophysical $S$ factor of the $^{3}{\rm He}(\alpha,
\gamma)^{7}{\rm Be}$ reaction}

\begin{figure}[htb]
\centerline{\includegraphics[width=16.8cm]{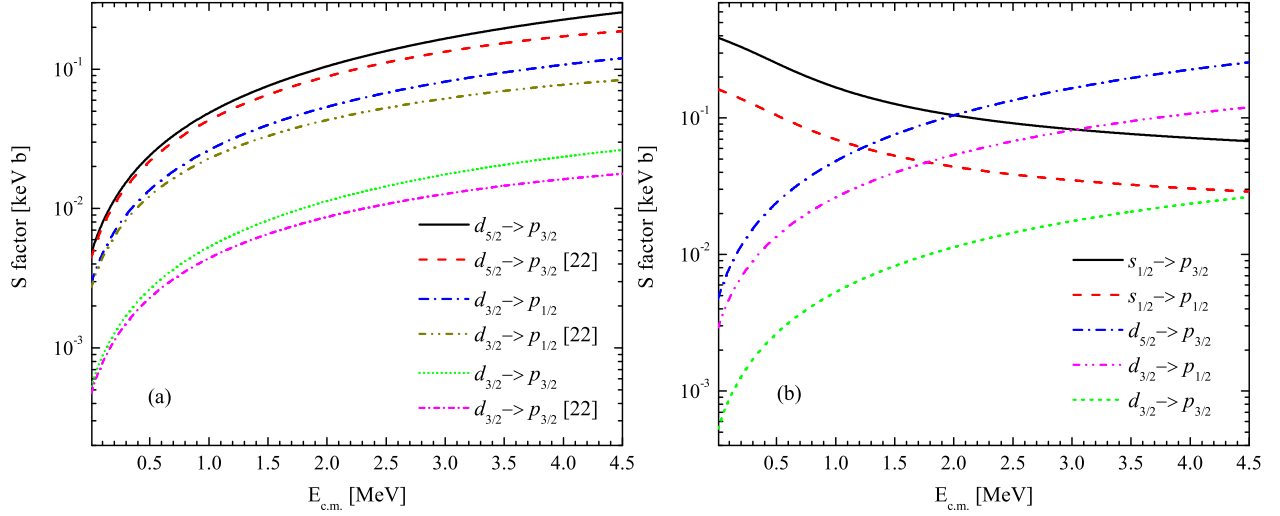}}
 \caption{Partial E1 astrophysical $S$ factors for the $^{3}{\rm He}(\alpha,
\gamma)^{7}{\rm Be}$ capture reaction calculated with the $V_D^n$ model potential in comparison with the results of Ref.
\cite{tur18_7Be}.}\label{f2}
\end{figure}

For the study of the $^{3}{\rm He}(\alpha, \gamma)^{7}{\rm Be}$
direct radiative capture process we first use the potential $V_D^n$.
Partial E1 astrophysical $S$ factors, estimated with
the $V_D^n$ potential are presented in Fig. \ref{f2}. Panel a compares the present results for the initial d-wave contribution with the corresponding ones obtained in Ref. \cite{tur18_7Be} using the potential model $V_D^a$. In panel b
the contributions from different initial $s$ and $d$ partial waves are
shown. As can be seen from the figure, the $d$-wave contribution
increases significantly at larger energies.

\begin{figure}[htb]
\centerline{\includegraphics[width=10cm]{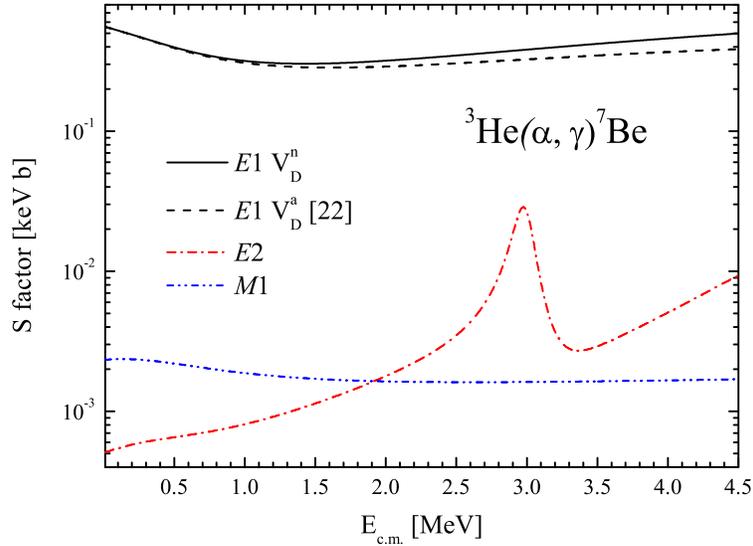}}
\caption{
E1, E2 and M1 components of the
astrophysical $S$ factors for the $^{3}{\rm He}(\alpha,
\gamma)^{7}{\rm Be}$ direct capture reaction calculated with the
model potential $V_D^n$. The corresponding E1 component from Ref.
\cite{tur18_7Be} is also shown.}
\label{f3}
\end{figure}

\begin{figure}[htb]
\centerline{\includegraphics[width=10cm]{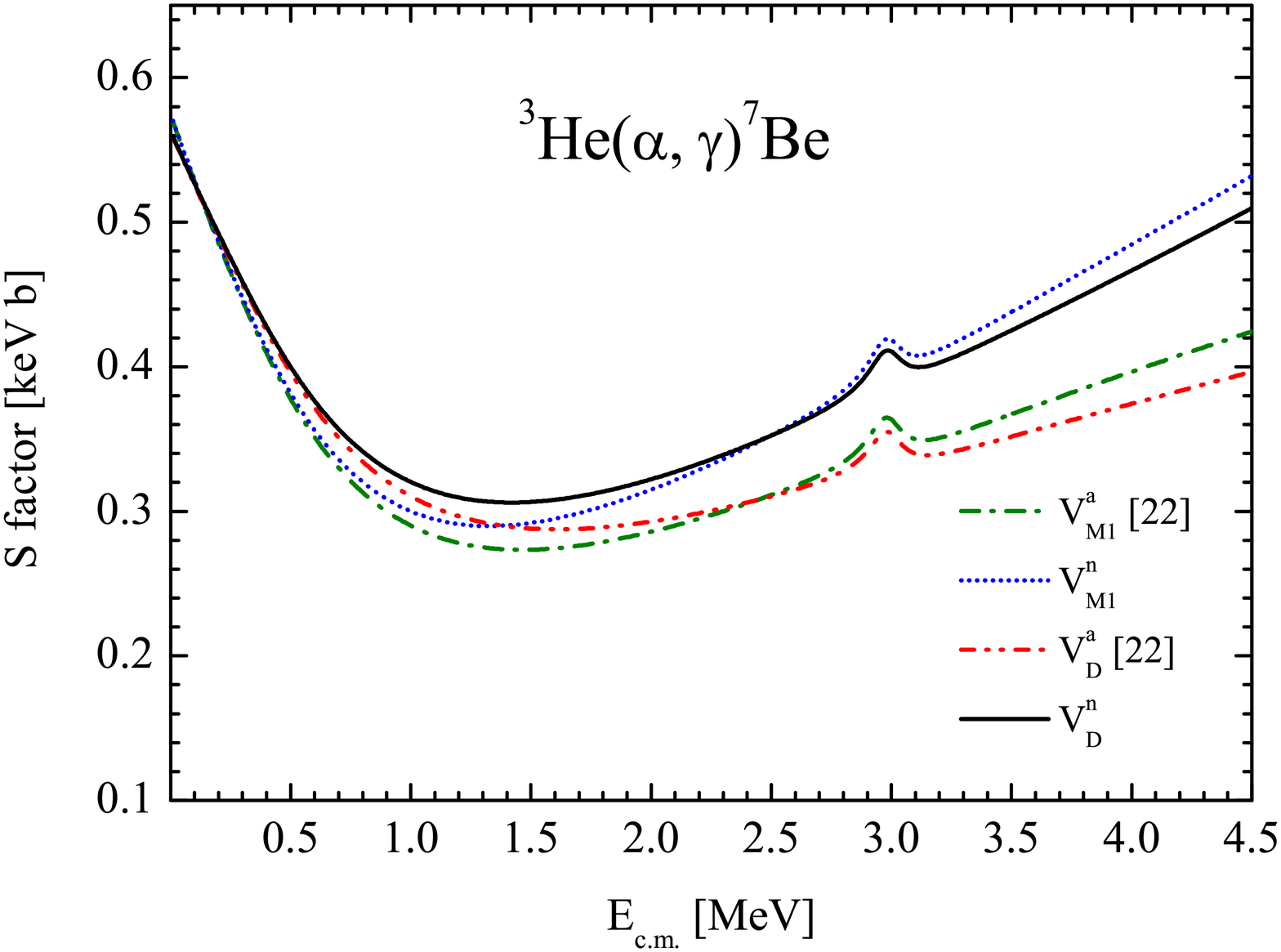}}
\caption{Astrophysical $S$ factor for the $^{3}{\rm He}(\alpha,
\gamma)^{7}{\rm Be}$ direct capture reaction calculated with
modified potentials $V_D^n$ and $V_{M1}^n$ in comparison with experimental data from Refs.
\cite{car12,sz13,takacs15,takacs18,sz19,nara04,luna06,bro07,luna07,leva09}
and the results of Ref. \cite{tur18_7Be}.}\label{f4}
\end{figure}

\begin{figure}[htb]
\centerline{\includegraphics[width=16.8cm]{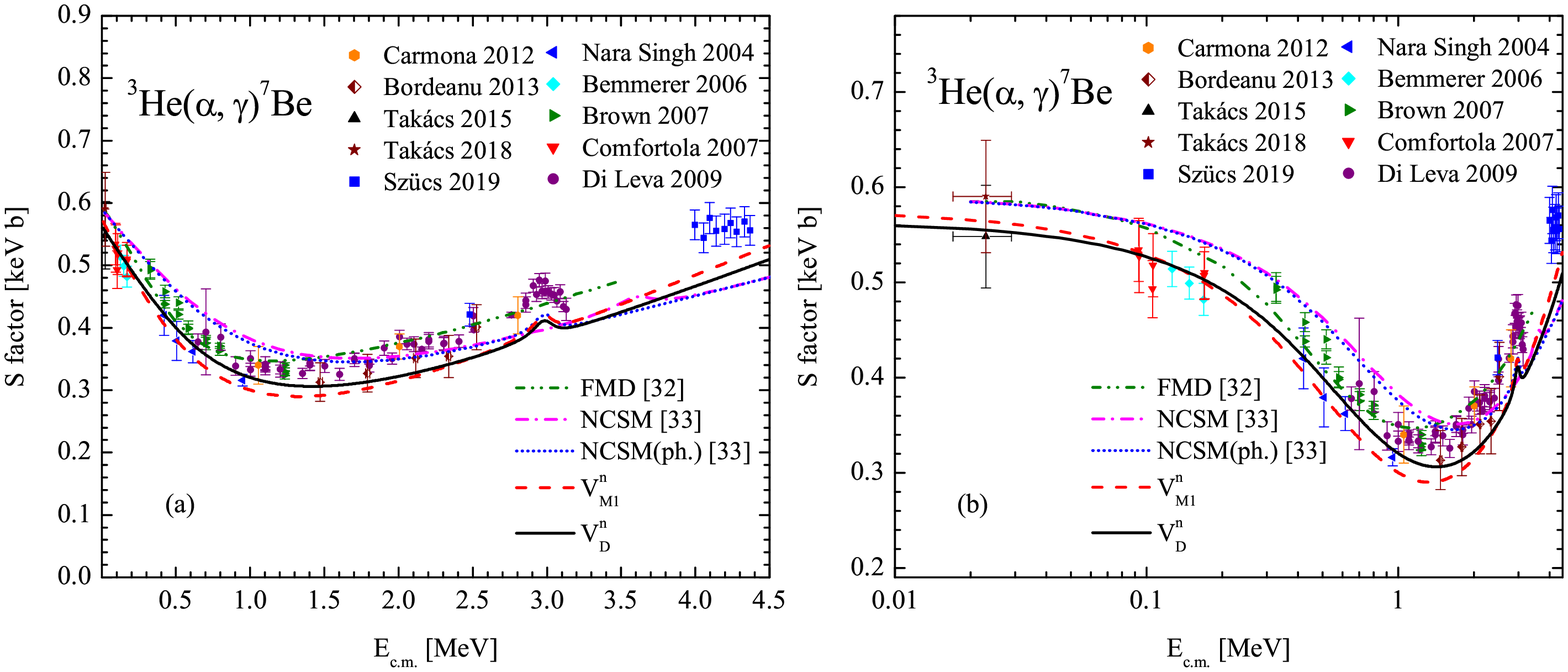}} \caption{(a)
Astrophysical $S$ factor for the $^{3}{\rm He}(\alpha, \gamma)^{7}{\rm
Be}$ direct capture reaction calculated with modified potential
models $V_D^n$ and $V_{M1}^n$ in comparison with
experimental data from Refs.
\cite{car12,sz13,takacs15,takacs18,sz19,nara04,luna06,bro07,luna07,leva09}
and \emph{ab-initio} calculations from Refs. \cite{neff11,doh16}.
Panel (b) highlights the low-energy region.}\label{f5}
\end{figure}

Contributions from the E1, E2 and M1 astrophysical
$S$ factors for the $^{3}{\rm He}(\alpha, \gamma)^{7}{\rm Be}$ direct
capture reaction calculated with the model potential $V_D^n$ are
presented in Fig. \ref{f3}. As can be seen from the figure, modification of
the potential in $d$ waves significantly  increases the astrophysical $S$ factor in comparison with the results of Ref.
\cite{tur18_7Be} at  energies above 0.5 MeV.

Figure \ref{f4} compares the astrophysical $S$ factors
calculated with modified potentials $V_D^n$ and $V_{M1}^n$ with
 experimental data from Refs.
\cite{car12,sz13,takacs15,takacs18,sz19,nara04,luna06,luna07,bro07,leva09}
and the results of Ref. \cite{tur18_7Be}. A substantial
improvement is achieved within the new models $V_D^n$ and
$V_{M1}^n$ at energies around and above the resonance energy.

In Fig. \ref{f5} the final results for the astrophysical $S$ factors
of the $^{3}{\rm He}(\alpha, \gamma)^{7}{\rm Be}$ direct capture
reaction are compared with the available data and results of
\emph{ab-initio} calculations from Refs. \cite{neff11,doh16}. As can
be seen from the figure, the potential models $V_D^n$ and $V_{M1}^n$
describe both absolute values and energy dependence of the
experimental data for the astrophysical $S$ factor in a wide energy
region from tens of keV to a few MeV.

\subsection{Astrophysical $S$ factor of the $^{3}{\rm H}(\alpha,
\gamma)^{7}{\rm Li}$ }

\begin{figure}[htb]
\centerline{\includegraphics[width=10cm]{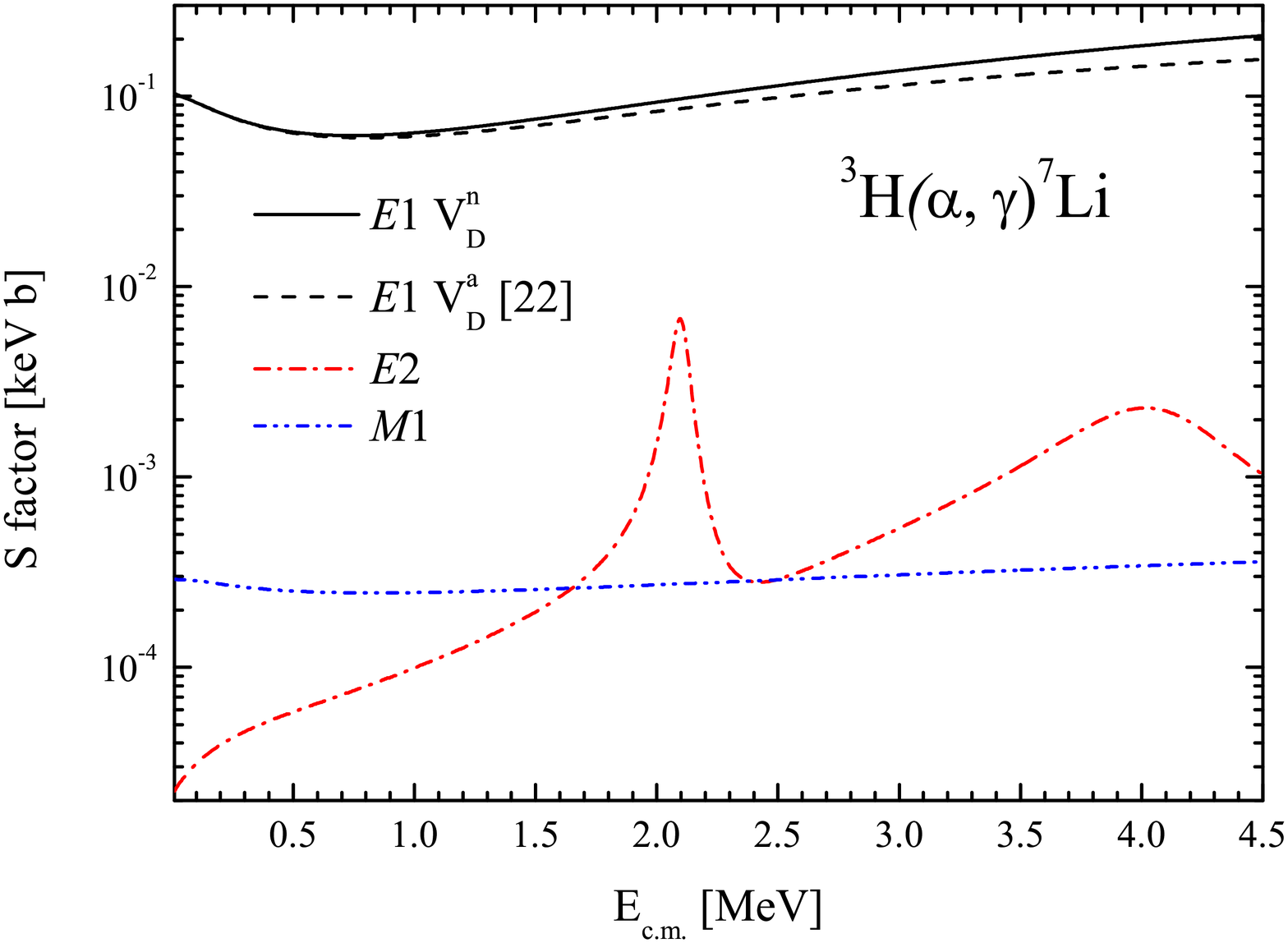}}
\caption{Comparison of contributions of the E1, E2 and M1
astrophysical $S$ factors for the $^{3}{\rm H}(\alpha, \gamma)^{7}{\rm
Li}$ direct capture reaction calculated with modified potential
model $V_D^n$ compared with the results of Ref.
\cite{tur18_7Be}.}\label{f6}
\end{figure}
\begin{figure}[htb]
\centerline{\includegraphics[width=10cm]{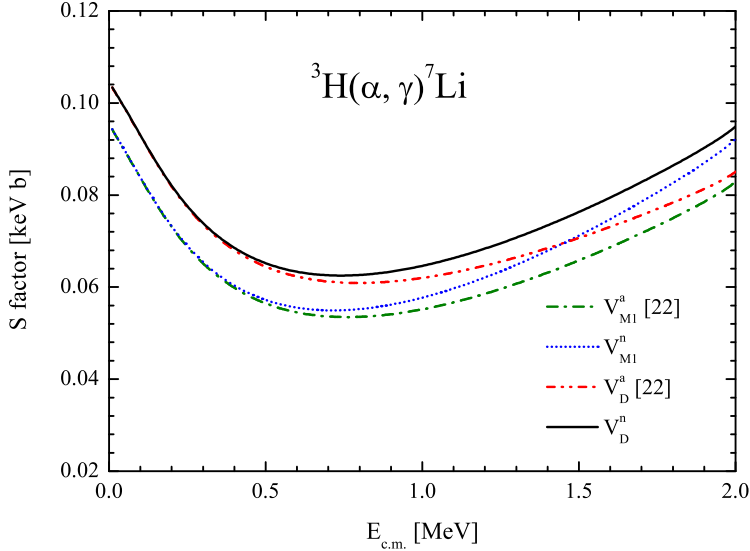}} \caption{
Astrophysical $S$ factor for the $^{3}{\rm H}(\alpha, \gamma)^{7}{\rm
Li}$ direct capture reaction calculated with modified potential
models $V_D^n$ and $V_{M1}^n$ in comparison with available
experimental data from Refs.
\cite{grif61,schr87,burz87,utsun90,brune94,tokim01,byst17} and the
results of Ref. \cite{tur18_7Be}.}\label{f7}
\end{figure}
\begin{figure}[htb]
\centerline{\includegraphics[width=10cm]{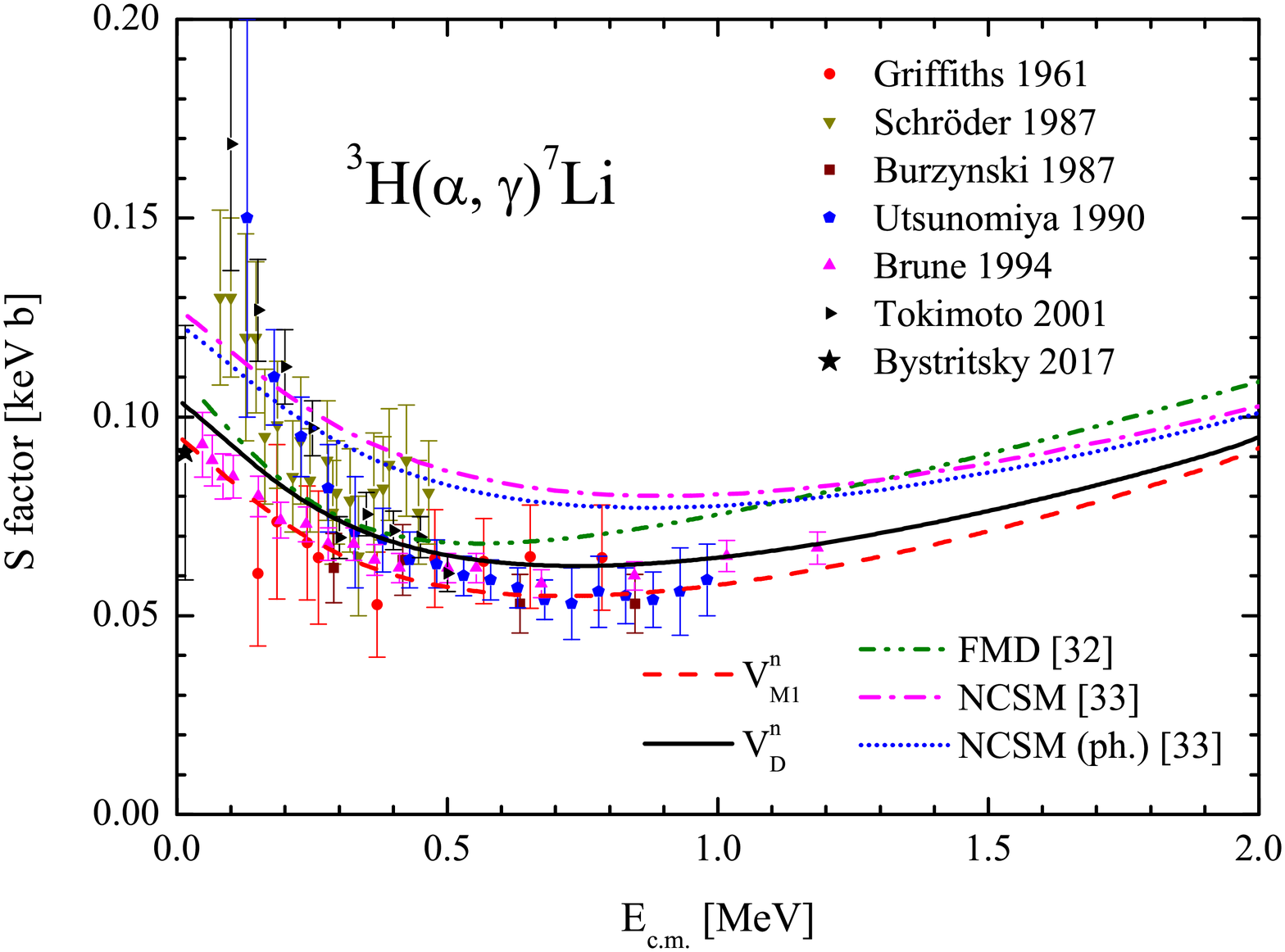}} \caption{(a)
Astrophysical $S$ factor for the $^{3}{\rm H}(\alpha, \gamma)^{7}{\rm
Li}$ direct capture reaction calculated with modified potential
models $V_D^n$ and $V_{M1}^n$ in comparison with available
experimental datafrom Refs.
\cite{grif61,schr87,burz87,utsun90,brune94,tokim01,byst17} and
\emph{ab-initio} calculations \cite{neff11,doh16}. Panel (b)
highlights the low-energy region.}\label{f8}
\end{figure}

As noted in the Introduction, the same model potentials $V_D^n$ and
$V_{M1}^n$ are applied for the study of the mirror capture reaction
$^{3}$H$(\alpha, \gamma)^{7}$Li. The Coulomb part of these
potentials, defined in Eq. (\ref{Coulomb}), is modified according to
the charge value of the $^3$H cluster, $Z$=1. As demonstrated in Fig.
\ref{f1} (panel b), the phase shifts in the $d_{3/2}$ and $d_{5/2}$ partial
waves are well described. The binding energies $E_b(3/2^-)$=2.467
MeV and $E_b(1/2^-)$=1.990 MeV of the bound states have been
reproduced in Ref. \cite{tur18_7Be}.

In Fig. \ref{f6} we compare the contributions of the $E$1, $E$2 and $M$1
astrophysical $S$ factors for the $^{3}{\rm H}(\alpha, \gamma)^{7}{\rm
Li}$ direct capture reaction calculated with the potentials
$V_D^n$ and $V_D^a$ from Ref. \cite{tur18_7Be}. As in the case of
$^{7}{\rm Be}$, the relative contribution of the E1 transition increases with the
energy in comparison with the results of Ref. \cite{tur18_7Be}.

Figure \ref{f7} presents the astrophysical $S$ factor for the $^{3}{\rm
H}(\alpha, \gamma)^{7}{\rm Li}$ direct capture reaction calculated
with modified potentials $V_D^n$ and $V_{M1}^n$ in comparison
with experimental data from Refs.
\cite{grif61,schr87,burz87,utsun90,brune94,tokim01,byst17} and the
results of Ref. \cite{tur18_7Be}. An increase of the astrophysical S
factor within the models $V_D^n$ and $V_{M1}^n$ is seen for energies
$E>0.5$ MeV. The best description of the data is obtained within the $V_{M1}^n$ model.

In Fig. \ref{f8} the astrophysical $S$ factors for the $^{3}{\rm
H}(\alpha, \gamma)^{7}{\rm Li}$ direct capture reaction calculated
with modified potential models $V_D^n$ and $V_{M1}^n$ are compared
with available experimental data and \emph{ab-initio} calculations.
As can be seen, the best description of the data for both
absolute value and energy dependence of the astrophysical $S$ factor
is obtained with the new potential models $V_D^n$ and $V_{M1}^n$. As
noted above, all the parameters of the model potentials have been
adjusted to the data for the $^{7}{\rm Be}$ nucleus. With that the
results for the astrophysical $S$ factor for the mirror $^{7}{\rm Li}$
nucleus are obtained without any fitting parameters.
Additionally, the same potentials describe the binding energies and
phase shifts for the mirror $^{7}{\rm Li}$ nucleus~\cite{tur18_7Be}.

\section{Reaction rates and primordial abundance of the $^7$Li element}
\subsection{Estimation of  reaction rates for the  $^{3}{\rm
He}(\alpha, \gamma)^{7}{\rm Be}$ process}

In Table \ref{ta1} estimated values for the reaction rate are given for
the $^{3}{\rm He}(\alpha, \gamma)^{7}{\rm Be}$ direct capture
process in the temperature interval $10^{6}$ K $\leq T \leq 10^{9}$
K ($ 0.001\leq T_{9} \leq 1 $). From the values presented in the table one can conclude that the
numerical results for the models $V_D^n$ and $V_{M1}^n$ are in a
good agreement with those obtained using the models $V_{D}^{a}$ and
$V_{M1}^a$ \cite{tur19}, respectively.

\begin{table}[htb]
\caption{Theoretical estimates of the reaction rates for the direct
$^{3}{\rm He}(\alpha, \gamma)^{7}{\rm Be}$ capture process in the
temperature interval $10^{6}$ K $\leq T \leq 10^{9}$ K ($ 0.001\leq
T_{9} \leq 1 $)} {\begin{tabular}{@{}cccccc@{}} \toprule $T_{9}$ &
$V_{M1}^{n}$ & $V_{D}^{n}$ & $T_{9}$& $V_{M1}^{n}$ & $V_{D}^{n}$\\
\colrule
0.001 & $9.554\times10^{-48}$ & $9.367\times10^{-48}$ & 0.070 & $9.755\times10^{-7}$ & $9.662\times10^{-7}$\\
0.002 & $1.949\times10^{-36}$ & $1.911\times10^{-36}$ & 0.080 & $3.450\times10^{-6}$ & $3.421\times10^{-6}$\\
0.003 & $5.896\times10^{-31}$ & $5.784\times10^{-31}$ & 0.090 & $1.001\times10^{-5}$ & $9.932\times10^{-6}$\\
0.004 & $1.677\times10^{-27}$ & $1.645\times10^{-27}$ & 0.100 & $2.498\times10^{-5}$ & $2.481\times10^{-5}$\\
0.005 & $4.775\times10^{-25}$ & $4.687\times10^{-25}$ & 0.110 & $5.545\times10^{-5}$ & $5.514\times10^{-5}$\\
0.006 & $3.547\times10^{-23}$ & $3.482\times10^{-23}$ & 0.120 & $1.121\times10^{-4}$ & $1.116\times10^{-4}$\\
0.007 & $1.104\times10^{-21}$ & $1.084\times10^{-21}$ & 0.130 & $2.102\times10^{-4}$ & $2.093\times10^{-4}$\\
0.008 & $1.877\times10^{-20}$ & $1.844\times10^{-20}$ & 0.140 & $3.699\times10^{-4}$ & $3.687\times10^{-4}$\\
0.009 & $2.057\times10^{-19}$ & $2.021\times10^{-19}$ & 0.150 & $6.175\times10^{-4}$ & $6.161\times10^{-4}$\\
0.010 & $1.615\times10^{-18}$ & $1.586\times10^{-18}$ & 0.160 & $9.856\times10^{-4}$ & $9.842\times10^{-4}$\\
0.011 & $9.773\times10^{-18}$ & $9.604\times10^{-18}$ & 0.180 & $2.249\times10^{-3}$ & $2.249\times10^{-3}$\\
0.012 & $4.805\times10^{-17}$ & $4.723\times10^{-17}$ & 0.200 & $4.562\times10^{-3}$ & $4.569\times10^{-3}$\\
0.013 & $1.995\times10^{-16}$ & $1.961\times10^{-16}$ & 0.250 & $1.864\times10^{-2}$ & $1.874\times10^{-2}$\\
0.014 & $7.196\times10^{-16}$ & $7.076\times10^{-16}$ & 0.300 & $5.401\times10^{-2}$ & $5.446\times10^{-2}$\\
0.015 & $2.307\times10^{-15}$ & $2.269\times10^{-15}$ & 0.350 & $1.254\times10^{-1}$ & $1.268\times10^{-1}$\\
0.016 & $6.694\times10^{-15}$ & $6.584\times10^{-15}$ & 0.400 & $2.496\times10^{-1}$ & $2.531\times10^{-1}$\\
0.018 & $4.400\times10^{-14}$ & $4.329\times10^{-14}$ & 0.450 & $4.447\times10^{-1}$ & $4.522\times10^{-1}$\\
0.020 & $2.224\times10^{-13}$ & $2.189\times10^{-13}$ & 0.500 & $7.286\times10^{-1}$ & $7.426\times10^{-1}$\\
0.025 & $5.683\times10^{-12}$ & $5.598\times10^{-12}$ & 0.600 & $1.630\times10^{0}$ & $1.668\times10^{0}$\\
0.030 & $6.692\times10^{-11}$ & $6.597\times10^{-11}$ & 0.700 & $3.072\times10^{0}$ & $3.157\times10^{0}$\\
0.040 & $2.408\times10^{-9}$ & $2.377\times10^{-9}$ & 0.800 & $5.150\times10^{0}$ & $5.312\times10^{0}$\\
0.050 & $3.049\times10^{-8}$ & $3.014\times10^{-8}$ & 0.900 & $7.929\times10^{0}$ & $8.204\times10^{0}$\\
0.060 & $2.100\times10^{-7}$ & $2.078\times10^{-7}$ & 1.000 & $1.145\times10^{1}$ & $1.189\times10^{1}$\\
\botrule
\end{tabular} \label{ta1}}
\end{table}

\begin{table}[htb]
\caption{Fitted values of the coefficients of analytical
approximation for the direct capture reaction $^{3}{\rm He}(\alpha,
\gamma)^{7}{\rm Be}$} {\begin{tabular}{@{}cccccccc@{}} \toprule
\textrm{Model} &
$p_0$ & $p_1$ & $p_2$ & $p_3$ & $p_4$ & $p_5$ & $p_6$ \\
\colrule

$V_{M1}^{n}$ & $2.697\times10^{6}$ & 8.105 & -26.574 & 42.958 & -35.272 &11.347 & 446.257 \\
$V_{D}^{n}$  & $2.636\times10^{6}$ & 8.155 & -26.704 & 43.602 & -35.987 &11.595 & 465.678 \\

\botrule
\end{tabular} \label{ta2}}
\end{table}

\begin{figure}[htb]
\centerline{\includegraphics[width=10.7cm]{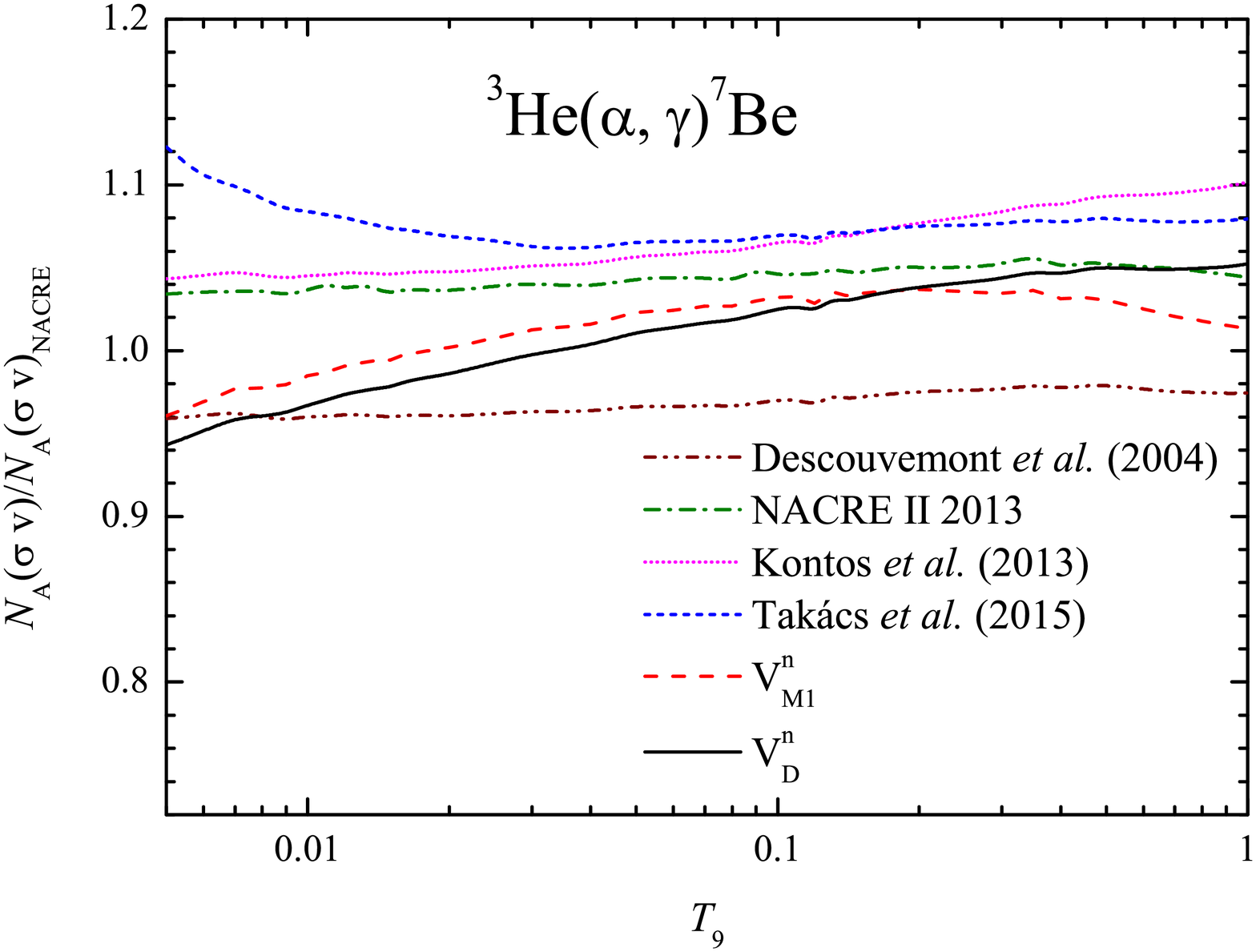}} \vspace*{8pt}
\caption{Reaction rates of the $^{3}{\rm He}(\alpha, \gamma)^{7}{\rm
Be}$ direct capture process normalized to the NACRE 1999
experimental data in comparison with results from
Refs.\cite{takacs15,kontos13,desc04} and more recent NACRE II 2013 data
\cite{nacre13}.\label{f9}}
\end{figure}

In Fig. \ref{f9} we present estimated reaction rates for the
direct $^{3}{\rm He}(\alpha, \gamma)^{7}{\rm Be}$ capture process
within the modified potential models $V_D^n$ and $V_{M1}^n$,
 normalized to the standard NACRE 1999 experimental data \cite{nacre99}.
 For comparison we also display the lines corresponding to the results of
Refs. \cite{takacs15,kontos13,desc04} and more recent NACRE II 2013 data
\cite{nacre13}. As can be seen from the figure, the potential model
results lye between the lines for the microscopic R-matrix approach
from Ref.\cite{desc04} and the NACRE II data. Other models
\cite{takacs15,kontos13} overestimate the NACRE II data.

In order to estimate the primordial abundance of the $^7$Li element
the well known PArthENoPE \cite{Pisanti08} public code is employed. It
 operates with an analytical form of the reaction rate
dependence on the temperature $T_9$. For this reason the theoretical
reaction rates from Table \ref{ta1} are approximated (within an uncertainty of 0.971\%
for the $V_{D}^{n}$ and 0.582\% for the $V_{M1}^{n}$) by
the analytical form
\begin{eqnarray}
\label{analitik}
 N_{A}(\sigma v)&=&p_0 T_{9}^{-2/3} \exp ( -C_{0} T_{9}^{-1/3}) \times ( 1 + p_1 T_9^{1/3} + p_2
 T_{9}^{2/3}+\\\nonumber
 &&+ p_3 T_{9}+p_4 T_{9}^{4/3}+p_5 T_{9}^{5/3}) + p_6 T_{9}^{-3/2} \exp (-C_{01} T_9^{-1}).
\end{eqnarray}
The coefficients of the
analytical polynomial approximation of the $^{3}{\rm He}(\alpha,
\gamma)^{7}{\rm Be}$ reaction rates estimated within the potential
models $V_{M1}^{n}$ and $V_{D}^{n}$ are given in Table \ref{ta2} in
the temperature interval $0.001\leq T_{9} \leq 1 $. In addition, for this
process the other coefficients are $C_{0}=12.813$ and $C_{01}=15.889$.

On the basis of the theoretical reaction rates and with the help of
the PArthENoPE \cite{Pisanti08} code we have estimated a
contribution from the $^{3}{\rm He}(\alpha, \gamma)^{7}{\rm Be}$
direct capture reaction to the primordial abundance of the $^7$Li
element. If we adopt the Planck 2015 best fit for the baryon density
parameter $\Omega_b h^2=0.02229^{+0.00029}_{-0.00027}$ \cite{ade16}
and the neutron life time $\tau_n=880.2 \pm 1.0$ s \cite{olive14},
for the $^7$Li/H abundance ratio we have an estimate $(4.930 \pm
0.129)\times 10^{-10}$ within potential model  $V_D^n$ and the
estimate $(4.842 \pm 0.126)\times 10^{-10}$ within the model
$V_{M1}^n$ which agree well, within 2\%, to be specific. As discussed below, these
numbers barely change the $^7$Li/H abundance ratio if the contribution from the
$^{3}{\rm H}(\alpha, \gamma)^{7}{\rm Li}$ direct capture reaction is included.

\subsection{Estimation of  reaction rates for the $^{3}{\rm
H}(\alpha, \gamma)^{7}{\rm Li}$ direct capture process}

\begin{table}[htb]
\caption{Theoretical estimates of the reaction rates for the
$^{3}{\rm H}(\alpha, \gamma)^{7}{\rm Li}$ direct capture process in
the temperature interval $10^{6}$ K $\leq T \leq 10^{9}$ K ($
0.001\leq T_{9} \leq 1 $)} {\begin{tabular}{@{}cccccc@{}} \toprule
$T_{9}$ &
$V_{M1}^{n}$ & $V_{D}^{n}$ & $T_{9}$& $V_{M1}^{n}$ & $V_{D}^{n}$\\
\colrule
0.001 & $5.595\times10^{-28}$ & $6.130\times10^{-28}$ & 0.070 & $1.326\times10^{-2}$ & $1.461\times10^{-2}$\\
0.002 & $6.285\times10^{-21}$ & $6.887\times10^{-21}$ & 0.080 & $2.839\times10^{-2}$ & $3.130\times10^{-2}$\\
0.003 & $1.613\times10^{-17}$ & $1.767\times10^{-17}$ & 0.090 & $5.384\times10^{-2}$ & $5.939\times10^{-2}$\\
0.004 & $2.252\times10^{-15}$ & $2.468\times10^{-15}$ & 0.100 & $9.318\times10^{-2}$ & $1.028\times10^{-1}$\\
0.005 & $7.497\times10^{-14}$ & $8.219\times10^{-14}$ & 0.110 & $1.502\times10^{-1}$ & $1.658\times10^{-1}$\\
0.006 & $1.081\times10^{-12}$ & $1.185\times10^{-12}$ & 0.120 & $2.286\times10^{-1}$ & $2.525\times10^{-1}$\\
0.007 & $9.075\times10^{-12}$ & $9.951\times10^{-12}$ & 0.130 & $3.324\times10^{-1}$ & $3.673\times10^{-1}$\\
0.008 & $5.234\times10^{-11}$ & $5.740\times10^{-11}$ & 0.140 & $4.651\times10^{-1}$ & $5.142\times10^{-1}$\\
0.009 & $2.297\times10^{-10}$ & $2.519\times10^{-10}$ & 0.150 & $6.304\times10^{-1}$ & $6.973\times10^{-1}$\\
0.010 & $8.194\times10^{-10}$ & $8.989\times10^{-10}$ & 0.160 & $8.316\times10^{-1}$ & $9.201\times10^{-1}$\\
0.011 & $2.488\times10^{-9}$ & $2.729\times10^{-9}$ & 0.180 & $1.353\times10^{0}$ & $1.499\times10^{0}$\\
0.012 & $6.641\times10^{-9}$ & $7.286\times10^{-9}$ & 0.200 & $2.052\times10^{0}$ & $2.274\times10^{0}$\\
0.013 & $1.596\times10^{-8}$ & $1.751\times10^{-8}$ & 0.250 & $4.677\times10^{0}$ & $5.191\times10^{0}$\\
0.014 & $3.516\times10^{-8}$ & $3.858\times10^{-8}$ & 0.300 & $8.672\times10^{0}$ & $9.640\times10^{0}$\\
0.015 & $7.201\times10^{-8}$ & $7.903\times10^{-8}$ & 0.350 & $1.409\times10^{1}$ & $1.568\times10^{1}$\\
0.016 & $1.386\times10^{-7}$ & $1.521\times10^{-7}$ & 0.400 & $2.091\times10^{1}$ & $2.330\times10^{1}$\\
0.018 & $4.408\times10^{-7}$ & $4.839\times10^{-7}$ & 0.450 & $2.905\times10^{1}$ & $3.242\times10^{1}$\\
0.020 & $1.191\times10^{-6}$ & $1.308\times10^{-6}$ & 0.500 & $3.843\times10^{1}$ & $4.293\times10^{1}$\\
0.025 & $8.679\times10^{-6}$ & $9.534\times10^{-6}$ & 0.600 & $6.045\times10^{1}$ & $6.767\times10^{1}$\\
0.030 & $3.920\times10^{-5}$ & $4.308\times10^{-5}$ & 0.700 & $8.615\times10^{1}$ & $9.661\times10^{1}$\\
0.040 & $3.484\times10^{-4}$ & $3.832\times10^{-4}$ & 0.800 & $1.148\times10^{2}$ & $1.289\times10^{2}$\\
0.050 & $1.629\times10^{-3}$ & $1.793\times10^{-3}$ & 0.900 & $1.457\times10^{2}$ & $1.638\times10^{2}$\\
0.060 & $5.244\times10^{-3}$ & $5.775\times10^{-3}$ & 1.000 & $1.783\times10^{2}$ & $2.008\times10^{2}$\\
\botrule
\end{tabular} \label{ta3}}
\end{table}

\begin{table}[htb]
\caption{Fitted values of the coefficients of analytical
approximation for the $^{3}{\rm H}(\alpha, \gamma)^{7}{\rm Li}$
direct capture reaction} {\begin{tabular}{@{}cccccccc@{}} \toprule
\textrm{Model} &
$p_0$ & $p_1$ & $p_2$ & $p_3$ & $p_4$ & $p_5$ & $p_6$ \\
\colrule

$V_{M1}^{n}$ & $4.948\times10^{5}$ & 4.053 & -13.252 & 21.105 & -17.624 &5.868 & 47.365 \\
$V_{D}^{n}$  & $5.422\times10^{5}$ & 4.042 & -13.159 & 21.080 & -17.681 &5.899 & 53.267 \\

\botrule
\end{tabular} \label{ta4}}
\end{table}

\begin{figure}[htb]
\centerline{\includegraphics[width=10cm]{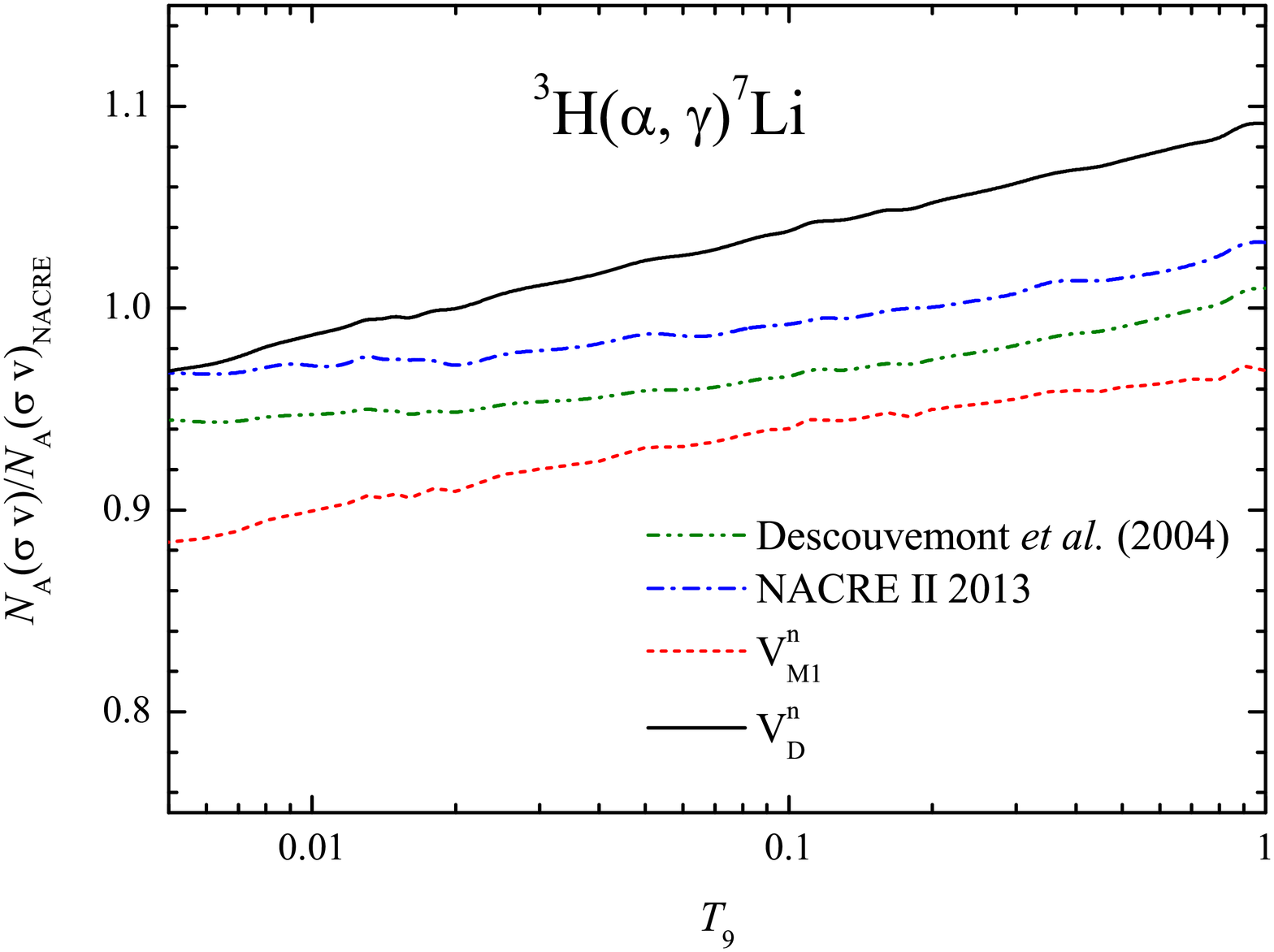}} \vspace*{8pt}
\caption{Reaction rates of the direct $^{3}{\rm H}(\alpha,
\gamma)^{7}{\rm Li}$ capture process normalized to the NACRE 1999
experimental data \cite{nacre99} in comparison with the results of
Ref.\cite{desc04} and new NACRE II 2013 data \cite{nacre13}
\label{f10}}
\end{figure}

In Table \ref{ta3} we give theoretical estimations for the $^{3}{\rm
H}(\alpha, \gamma)^{7}{\rm Li}$ direct capture reaction rates in the temperature
interval $10^{6}$ K $\leq T \leq 10^{9}$ K ($ 0.001\leq T_{9} \leq 1
$) calculated with the same modified potential models $V_{M1}^{n}$
and $V_{D}^{n}$ which have been used for the $^{3}{\rm He}(\alpha,
\gamma)^{7}{\rm Be}$ process.
Figure \ref{f10} displays
these results
normalized
to the standard NACRE 1999 experimental data \cite{nacre99}. For the
comparison we also display the lines corresponding to the results of
the microscopic R-matrix method \cite{desc04} and new NACRE II
2013 data \cite{nacre13}.

The coefficients of the analytical polynomial approximation of the
$^{3}{\rm H}(\alpha, \gamma)^{7}{\rm Li}$ reaction rates estimated
within the potential models $V_{M1}^{n}$ and $V_{D}^{n}$ are given
in Table \ref{ta4} in the temperature interval $0.001\leq T_{9}
\leq 1$. The remaining coefficients are $C_{0}=8.072$ and
$C_{01}=3.689$. In this case, the analytical formula (\ref{analitik})
with the parameter values from Table \ref{ta4} reproduces the
theoretical reaction rates from Table \ref{ta3} (within an uncertainty 0.599\% for
 $V_{D}^{n}$ and 0.647\% for $V_{M1}^{n}$).

Now  including the obtained theoretical reaction rates for both  $^{3}{\rm
He}(\alpha, \gamma)^{7}{\rm Be}$ and $^{3}{\rm
H}(\alpha,\gamma)^{7}{\rm Li}$ capture processes into the nuclear
reaction network with the help of the PArthENoPE \cite{Pisanti08}
code, we can evaluate the primordial abundance of the $^7$Li
element. Adopting the aforementioned values of the baryon density and
the neutron life time, for the $^7$Li/H abundance ratio we have an
estimate $(4.936 \pm 0.129)\times 10^{-10}$ within the model
$V_{D}^{n}$, while the model $V_{M1}^{n}$ yields $(4.835 \pm
0.127)\times 10^{-10}$ \cite{tur19}. These numbers are slightly
different than the corresponding estimates based exclusively on the $^{3}{\rm
He}(\alpha, \gamma)^{7}{\rm Be}$  process.

\section{Conclusions}

The astrophysical $^{3}{\rm He}(\alpha, \gamma)^{7}{\rm Be}$ and
$^{3}{\rm H}(\alpha, \gamma)^{7}{\rm Li}$ direct capture reactions
have been studied in an updated two-body potential model. The
parameters of the central potentials of a simple Gaussian form have
been adjusted to reproduce the $\alpha+^3${\rm He} phase shifts in
the $s$, $p$, $d$ and $f$ partial waves and the binding energies of
the $^7$Be ground 3/2$^-$ and first excited 1/2$^-$ states. At the
same time, properties of the mirror $^7$Li nucleus, phase shifts in
the partial waves and the binding energies of the ground 3/2$^-$ and
first excited 1/2$^-$ states are reproduced without any additional
adjustment parameters.

It is found that due to the dominance of the E1 transition in the
capture processes, there is a possibility to adjust the parameters of
the potential in the initial $s$- and $d$-waves in order to optimize
the description of the astrophysical $S$ factor at low and intermediate
energy regions, respectively.

In conclusion, the potential models $V_{M1}^n$, $V_{D}^n$ have been
suggested for the description of the $\alpha+^3H$ and $\alpha+^3He$
interactions. These models reproduce spectroscopic properties and
phase shifts of both $^{7}{\rm Be}$ and $ ^{7}{\rm Be}$ nuclei. They
describe well the experimental data for the astrophysical $S$ factor
of the capture process $^{3}{\rm He}(\alpha, \gamma)^{7}$Be in a wide
energy region, extending to 4.5 MeV. This includes the new data of the LUNA
collaboration around 100 keV and the latest data at the Gamov peak
obtained on the basis of the observed neutrino fluxes from the Sun,
$S_{34}$(23$^{+6}_{-5}$ keV)=0.548$\pm$0.054 keV b. The same
potentials describe the astrophysical $S$ factor for the mirror
capture reaction $^{3}{\rm H}(\alpha, \gamma)^{7}{\rm Li}$ with a
good accuracy.

The calculated values of the astrophysical $S$ factors and reaction
rates for the $^{3}{\rm He}(\alpha, \gamma)^{7}{\rm Be}$ and
$^{3}{\rm H}(\alpha, \gamma)^{7}{\rm Li}$ direct capture reactions
are in good agreement with the results of microscopic models and
\emph{ab-initio} calculations. For the primordial abundance of the
$^7$Li element an estimate $(4.89 \pm 0.18)\times 10^{-10}$ have
been obtained. This result is within the range of the standard BBN model
estimates.

\section*{Acknowledgements}
We thank J. Dohet-Eraly for providing us with the results of Ref.
\cite{doh16} for the astrophysical $S$ factor in a tabulated form.
A.S.K. acknowledges the support from the Australian
Research Council.


\begin{thebibliography}{99}
\bibitem{sbor10} L. Sbordone, P. Bonifacio, E. Caffau, Astron.
and Astrophys. {\bf 522} (2010) A26
\bibitem{cyb16} R.H. Cyburt, B.D. Fields, K.A. Olive and T-H. Yeh, Reviews of Modern Physics {\bf 88}, 1 (2016)
\bibitem{luna17} LUNA Collaboration (D.~Trezzi, {\it et~al.}) {\it  Astropart. Phys.} {\bf 89} 57 (2017)
\bibitem{asp06} M. Asplund, et al., Astrophys. J. {\bf 644} (2006) 229.
\bibitem{WS2019}
Proc. of Inter Conf. "Lithium in the universe: to be or not to be? Monte Porzio Catone, November 18-22, 2019 Editors: G. Cescutti, A. Korn and P. Ventura".
MEMORIE DELLA SOCIETA ASTRONOMICA ITALIANA Vol. 91 (2020), pp. 1-179.
\bibitem{tur16} E.M. Tursunov, A.S. Kadyrov, S.A. Turakulov, I. Bray, Phys. Rev. C
{\bf 94} (2016) 015801.
\bibitem{bt18} D. Baye, E.M. Tursunov, J. Phys. G, Nucl. Part. Phys. {\bf 45} (2018)
085102.
\bibitem{tur18} E.M. Tursunov, S.A. Turakulov, A.S. Kadyrov,
I. Bray, Phys. Rev. C {\bf 98} (2018) 055803.
\bibitem{tur20} E.M. Tursunov, S.A. Turakulov, A.S. Kadyrov, Nucl. Phys. {\bf A1000} (2020) 121884
\bibitem{luna14} LUNA Collaboration (M.~Anders, {\it et~al.}) {\it  Phys.~ Rev.~ Lett.}  {\bf 113}, 042501 (2014).
\bibitem{tur15} E.M. Tursunov, S.A. Turakulov, P. Descouvemont, Phys. At. Nucl.
{\bf 78} (2015) 193.
\bibitem{adel11} E. G. Adelberger  \emph{et al.},  Rev. Mod. Phys. {\bf 83}, 195 (2011).
\bibitem{fields11} B. D. Fields, Ann. Rev. Nucl. Particle Sci. {\bf 61} (2011) 47.
\bibitem{serene13} A. Serenelli, C. Pe\~{n}a-Garay, and W. C. Haxton, Phys. Rev. D {\bf 87},
043001 (2013).
\bibitem{coc17} A. Coc and E. Vangioni, Int. J. Mod. Phys. E {\bf 26}, 1741002 (2017).
\bibitem{luna06} D. Bemmerer, F. Confortola, H.Costantini, A. Formicola, Gy. Gy\"{u}rky, \emph{et al.}, Phys. Rev.
Lett. {\bf 97}, 122502 (2006).
\bibitem{luna07} F. Confortola, D. Bemmerer, H. Costantini, A. Formicola, Gy. Gy\"{u}rky, \emph{et al.}, Phys. Rev. C {\bf 75},
065803 (2007).
\bibitem{takacs15} M.P. Tak\'{a}cs, D. Bemmerer, T. Sz\"{u}cs, and K.Zuber, Phys. Rev. D {\bf 91}, 123526 (2015).
\bibitem{takacs18} M.P. Tak\'{a}cs, D. Bemmerer, A.R.Junghans, K.Zuber, Nucl. Phys. A {\bf 970}, 78 (2018).
\bibitem{sz13} C. Bordeanu, Gy. Gy\"{u}rky, Z. Hal\'{a}sz, T. Sz\"{u}cs, G. G. Kiss,  Z. Elekes, J. Farkas,
 Zs. F\"{u}l\"{o}p, and E. Somorjai, Nucl. Phys. A {\bf 908}, 1 (2013).
\bibitem{sz19} T. Sz\"{u}cs, G.G. Kiss, Gy. Gy\"{u}rky, Z. Hal\'{a}sz, T. N. Szegedi, Zs. F\"{u}l\"{o}p, Phys. Rev. C {\bf 99}, 055804 (2019).
\bibitem{tur18_7Be} E.M. Tursunov, S.A. Turakulov and A.S. Kadyrov,  Phys. Rev. C {\bf 97}, 035802 (2018).
\bibitem{dub10} S.B.~Dubovichenko, Physics of Atomic Nuclei, {\bf 73}, 1526 (2010).
\bibitem{mohr09} P. Mohr, Phys. Rev. C {\bf 79}, 065804 (2009).
\bibitem{mason09} A. Mason, R. Chatterjee, L. Fortunato, and A. Vitturi, Eur. Phys. J.  A {\bf 39}, 107 (2009).
\bibitem{kajino87} T. Kajino, Astrophysics J. {\bf 319}, 531 (1987).
\bibitem{vasil12} V. S. Vasilevsky, A. V. Nesterov, and T. P. Kovalenko, Physics of Atomic Nuclei, {\bf 75}, 818 (2012).
\bibitem{sol14} A.S.~ Solovyev, S.Yu.~ Igashov, Yu.M.~Tchuvilsky, J. Phys. CS {\bf 569},
0122020 (2014).
\bibitem{sol19} A.S.~ Solovyev, S.Yu.~ Igashov, Phys.
Rev. C {\bf 99}, 054618 (2019).
\bibitem{desc10} P. Descouvemont and D. Baye, Rep. Prog. Phys. {\bf 73}, 036301 (2010).
\bibitem{noll01} K.M. Nollett, Phys.Rev. C {\bf 63}, 054002 (2001).
\bibitem{neff11} T. Neff, Phys.Rev.Lett. {\bf 106}, 042502 (2011).
\bibitem{doh16} J. Dohet-Eraly, P. Navratil, S. Quaglioni, W. Horiuchi, G. Hupin and F. Raimondi, Phys.Lett. B {\bf 757}, 430 (2016).
\bibitem{vor19} M. Vorabbi,  P. Navratil, S. Quaglioni, G. Hupin, Phys.Rev. C {\bf 100}, 024304 (2019).
\bibitem{olim16} R.~Yarmukhamedov, O.R. Tojiboev, and S.V. Artemov, Nuovo Cimento C
39, 364 (2016).
\bibitem{nacre99} NACRE (C. Angulo {\it et al}.),  Nucl. Phys. A {\bf 656}, 3 (1999).
\bibitem{mukh16} A.M.~ Mukhamedzhanov, Shubhchintak, and C.A.~ Bertulani, Phys. Rev. C {\bf 93}, 045805 (2016).
\bibitem{spiger}R.J.~Spiger and T.A.~Tombrello, Phys.Rev. {\bf 163}, 964 (1967)
\bibitem{grif61} G. M. Griffiths, R. A. Morrow, P. J. Riley, and J. B.Warren, Can. J. Phys. {\bf 39}, 1397 (1961).
\bibitem{schr87} U. Schr\"{o}der, A. Redder, C. Rolfs, R. E. Azuma, L. Buchmann, C.
 Campbell, J. D. King, and T. R. Donoghue, Phys. Lett. B {\bf 192}, 55 (1987).
\bibitem{burz87} S. Burzy\'{n}ski, K. Czerski, A. Marcinkowski, and P. Zupranski, Nucl. Phys. A {\bf 473}, 179(1987).
\bibitem{utsun90} H. Utsunomiya, Y.-W.~Lui, \emph{et al.}, Phys.Rev.Lett. {\bf 65}, 847 (1990).
\bibitem{brune94} C. R. Brune, R.W. Kavanagh, and C. Rolfs, Phys. Rev. C {\bf 50}, 2205 (1994).
\bibitem{tokim01} Y. Tokimoto, H. Utsunomiya, \emph{et al.}, Phys. Rev. C {\bf 63},
035801 (2001).
\bibitem{byst17} V. M. Bystritsky, G. N. Dudkin, E. G.Emets, M. Filipowicz, A. R. Krylov, B. A. Nechaev, A. Nurkin,
 V. N. Padalko, A. V. Philippov, and A. B. Sadovsky, Phys. Part. Nucl. Lett. {\bf 14}, 560 (2017)
\bibitem{nara04} B. S. Nara Singh, M. Hass, Y. Nir-El, and G. Haquin, Phys. Rev. Lett. {\bf 93}, 262503 (2004).
\bibitem{bro07} T.A.D. Brown, C. Bordeanu, K. A. Snover, D. W. Storm, D. Melconian,
 A. L. Sallaska, S. K. L. Sjue, and S. Triambak, Phys. Rev. C {\bf 76}, 055801 (2007).
\bibitem{leva09} A. Di Leva, L. Gialanella, R. Kunz, D. Rogalla, D. Sch\"{u}rmann, \emph{et al.}, Phys. Rev. Lett. {\bf 102}, 232502 (2009).
\bibitem{car12} M. Carmona-Gallardo, B. S. Nara Singh, M. J. G. Borge, J. A. Briz, M. Cubero,
\emph{et al.}, Phys. Rev. C {\bf 86}, 032801 (2012).
\bibitem{tur19} S.A. Turakulov, E.M. Tursunov,  Int. J. Mod. Phys: Conf. Ser. {\bf 49}, 1960014 (2019).
\bibitem{kontos13} A. Kontos, E. Uberseder, R. deBoer, J. G\"orres, C. Akers, A. Best, M. Couder, M. Wiescher, Phys. Rev. C
{\bf 87}, 065804 (2013).
\bibitem{desc04} P. Descovemont, A.Adahchour, C. Angulo, A. Coc and E. Vangione-Flam, Atomic Data and
Nuclear Data Tables {\bf 88}, 203 (2004)
\bibitem{nacre13} NACRE II (Y.~Xu {\it et al}.), Nucl. Phys. A {\bf 918}, 61 (2013).
\bibitem{Pisanti08} O.~Pisanti, A.~Cirillo, S.~Esposito, F.~Iocco, G.~Mangano, G.~Miele, and P. D.~Serpico, Comput. Phys. Commun. {\bf 178}, 956 (2008).
\bibitem{ade16} Planck Collaboration (P. A. R.~Ade {\it et al}.), Astron. Astrophys.
{\bf 594}, A13 (2016).
\bibitem{olive14} M. Tanabashi {\it et al}. (Particle Data Group), Phys. Rev. {\bf D98}, 030001 (2018)
%
\bibitem{fraser} P. R. Fraser, K. Massen-Hane, A.S. Kadyrov, K. Amos, I. Bray, and L. Canton, Phys. Rev. C {\bf 96}, 014619 (2017).
\bibitem{boykin}W.R.~Boykin, S.D.~Baker, D.M.~Hardy, Nucl. Phys. A {\bf 195}, 241 (1972).
\bibitem{hardy}D.M.~Hardy, R.J.~Spiger, S.D.~Baker, Y.S.~Chen, T.A.~Tombrello, Nucl. Phys. A {\bf 195}, 250 (1972).

\end{thebibliography}
\end{document}